\documentclass[12pt]{article}

\usepackage{graphicx}
\usepackage{amssymb}
\begin{document}

\renewcommand{\arraystretch}{1.5}
\newcommand{\vecx}{\mbox{\boldmath $x$}}


%
\begin{center}
{\large\bf
Generalized information entropies in nonextensive 
quantum systems: The interpolation approach
} 
\end{center}

\begin{center}
Hideo Hasegawa
\footnote{hideohasegawa@goo.jp}
\end{center}

\begin{center}
{\it Department of Physics, Tokyo Gakugei University,  \\
Koganei, Tokyo 184-8501, Japan}
\end{center}
\begin{center}
({\today})
\end{center}
\thispagestyle{myheadings}

\begin{abstract}
We discuss the generalized von Neumann (Tsallis) entropy 
and the generalized Fisher information (GFI) in nonextensive quantum systems,
by using the interpolation approximation (IA) which has been shown to yield
good results for the quantal distributions within $O(q-1)$ and in high- 
and low-temperature limits, $q$ being the entropic index
[H. Hasegawa, Phys. Rev. E 80 (2009) 011126].
Three types of GFIs which have been proposed so far in the nonextensive
statistics, are discussed from the viewpoint of their metric properties 
and the Cram\'{e}r-Rao theorem.
Numerical calculations of the $q$- and temperature-dependent 
Tsallis entropy and GFIs are performed for the electron band model
and the Debye phonon model.
\end{abstract}

\vspace{0.5cm}
\vspace{1cm}

 

\newpage
\section{Introduction}

The Boltzmann-Gibbs-Shannon-von Neumann entropy and
the Fisher information play central roles as information
measures in classical and quantum statistics 
(for review see \cite{Frieden98}, relevant references therein).
The Boltzmann-Gibbs-Shannon-von Neumann entropy
represents a global measure of ignorance, while
the Fisher information expresses a loal measure of a positive 
amount of information \cite{Frieden98}.
The Fishser information has the two properties: (i) it
expresses the metric tensor for the neighboring points
in the Riemannian space spanned by probability distributions
or density matrices, and (ii) it provides the lower bound of unbiased 
estimation errors in the Cram\'{e}r-Rao theorem (CRT). 
The Fisher information has been employed for a study of
effecientcy of information transmission and its decodings. 

In the last decade, much progress has been made in the nonextensive
statistics initiated by Tsallis \cite{Tsallis88}, who proposed 
the generalized entropy (called the Tsallis entropy) defined by
\begin{eqnarray}
S_q 
&=& \frac{k_B}{q-1} \left[ 1 - \int p_q(x)^q \:dx \right] 
\hspace{1cm}\mbox{for classical case},
\label{eq:A1}\\ 
&=& \frac{k_B}{q-1} \left( 1 -Tr\: \hat{\rho}_q^q \right)
\hspace{2.5cm}\mbox{for quantum case}. 
\label{eq:A2}
\end{eqnarray}
Here $q$ denotes the entropic index, $k_B$ the Boltzmann constant,
$p_q(x)$ the probability distribution and $\hat{\rho}_q$ the density matrix.
The Tsallis entropy is a one-parameter generalization of
the Boltzmann-Gibbs-Shannon-von Neumann entropy, to which it reduces
in the limit of $q \rightarrow 1.0$.
The Tsallis entropy is non-additive in a sense that
for $p(A \cup B)=p(A)\:p(B)$ or
$\hat{\rho}(A \cup B)=\hat{\rho}(A) \otimes \hat{\rho}(A)$,
we obtain
\begin{eqnarray}
S_q(A \cup B) &=& S_q(A)+ S_q(B)+\frac{(1-q)}{k_B}S_q(A) S_q(B), \\
&\neq & S_q(A)+ S_q(B)
\hspace{2cm}\mbox{for $q\neq 1$}.
\end{eqnarray}
The Tsallis entropy is super-extesnsive (sub-extensive) for $q < 1$ ($q > 1$),
and $q-1$ expresses the degree of the nonextensivity. 
The nonextensive statistics has been widely applied to various subjects 
in physics, chemistry, information science, biology and economics \cite{Tsallis04}.

At the moment, three types of generalized Fisher informations (GFIs) 
have been proposed in the nonextensive statistics \cite{Masi06}-\cite{Hasegawa08},
\begin{eqnarray}
g_{\theta_n \theta_m} &=&
q  \;Tr \: \left[\hat{\rho}_q \left(\hat{\rho}_q^{-1} \frac{\partial \hat{\rho}_q}{\partial \theta_n} \right)
\left(\hat{\rho}_q^{-1} \frac{\partial \hat{\rho}_q}{\partial \theta_m} \right) \right],
\label{eq:A5} \\
G_{\theta_n \theta_m} &=&
Tr \: \left[\hat{P}_q \left(\hat{P}_q^{-1} \frac{\partial \hat{\rho}_q}{\partial \theta_n} \right)
\left(\hat{P}_q^{-1} \frac{\partial \hat{\rho}_q}{\partial \theta_m} \right) \right],
\label{eq:A6} \\
\tilde{g}_{\theta_n \theta_m} &=&
Tr \: \left[\hat{P}_q \left(\hat{P}_q^{-1} \frac{\partial \hat{P}_q}{\partial \theta_n} \right)
\left(\hat{P}_q^{-1} \frac{\partial \hat{P}_q}{\partial \theta_m} \right) \right],
\label{eq:A8} 
\end{eqnarray}
where $\hat{P}_q$ ($=\hat{\rho}_q^q/Tr \:\hat{\rho}_q^q$) expresses the escort density matrix
and $\theta_n$ a parameter specifying $\hat{\rho}_q$:
the classical case of Eqs. (\ref{eq:A5})-(\ref{eq:A8}) is obtainable 
if we read $\hat{\rho}_q \rightarrow p_q(x)$ and $Tr \rightarrow \int dx $
[see Eqs. (\ref{eq:K4})-(\ref{eq:K6})]. 
In the limit of $q \rightarrow 1.0$,
the three GFIs given by Eqs. (\ref{eq:A5}), (\ref{eq:A6}) and (\ref{eq:A8}) reduce to
the conventional expression,
\begin{eqnarray}
g_{\theta_n \theta_m} 
&=& Tr\: \left[ \hat{\rho}_1 
\left( \frac{\partial \ln \hat{\rho}_1}{\partial \theta_n} \right) 
\left( \frac{\partial \ln \hat{\rho}_1}{\partial \theta_m} \right)
\right].
\label{eq:A4}
\end{eqnarray}
The GFI in Eq. (\ref{eq:A5}) \cite{Masi06,Portesi07} which is derived
with the use of the generalized Kullback-Leibler divergence \cite{Abe03b,Abe04,Lanberti04},
expresses the metric tensor of the first properties (i) of the Fisher
information discussed above.
The GFI in Eq. (\ref{eq:A5}), however, is not applicable to 
the CRT for variance and covariance of physical quantities
averaged over the escort density matrix. 
The GFI given by Eq. (\ref{eq:A6}) \cite{Naudts04,Naudts05} 
preserves the second properties (ii) of the lower bound in the CRT,
although it does not have the metric properties (i).
The GFI given by Eq. (\ref{eq:A6}) \cite{Chimento00,Pennini04,Hasegawa08}
has both the properties (i) and (ii).
Details of the three GFIs and a comparison among them will be discussed
in this paper (Secs. III and V). 

The Tsallis entropy and generalized Fisher information (GFI) in nonextensive 
{\it classical} systems have been considerably studied in
\cite{Masi06,Naudts04,Chimento00,Pennini04,Hasegawa08}\cite{Tsallis95}-\cite{Abe03}.
In recent years, the generalized von Neumann 
({\it quantum} Tsallis) entropy has been investigated in
the bipartite spin-$1/2$ \cite{Abe02}, 
$1d$ spin-$1/2$ \cite{Caruso06}, $2d$ bosonic systems \cite{Caruso06}
and Hubbard dimers \cite{Hasegawa07}.
The purpose of this paper is to discuss the Tsallis entropy as well as the GFI
in nonextensive {\it quantum} systems.
Physical quantities in nonextensive quantum systems are
evaluated by the trace over the escort density matrix \cite{Martinez00}, 
which may be formally expressed in exact integral representations 
\cite{Rajagopal98,Lenzi99}. 
Their actual evaluations are, however, tedious and difficult
because they involve self-consistent calculations
of expectation values of the energy and number of particles.
This is the case in a calculation of the generalized Bose-Einstein and Fermi-Dirac
distributions (referred to $q$-BED and $q$-FDD, respectively)
in nonextensive quantum statistics.
Quite recently it has been pointed out that this difficulty may be
overcome when we adopt the interpolation approximation (IA)
which yields good results within the $O(q-1)$ and in the high- and
low-temperature limits \cite{Hasegawa09b}.
Indeed the $q$-BED and $q$-FDD in the IA are given in simple analytic 
expressions, which have been successfully
applied to nonextensive quantum systems such as black-body radiation,
Bose-Einstein condensation, BCS superconductivity and metallic
ferromagnetism \cite{Hasegawa09c}.
The three GFIs which are formally expressed in exact integral representations,
may be much simplified when we adopt the IA \cite{Hasegawa09b}.

The paper is organized as follows.
In Sec. II, we briefly explain the exact and interpolation approaches
to nonextensive quantum statistics \cite{Hasegawa09b}.
In Sec. III, the Tsallis entropy and the three GFIs in the IA are discussed.
Numerical calculations are reported in Sec. IV for the electron band model
and for the Debye phonon model.
Discussion and conclusion are presented in Sec. V, where
a comparison among the three GFIs is made. 
The CRT for the three GFIs in \cite{Chimento00,Pennini04,Hasegawa08,Naudts05} 
is discussed also for the $q$-Gaussian distribution in the nonextensive classical
statistics.

\section{Exact and interpolation approaches}

\subsection{Exact approach}

We first obtain the optimum density matrix $\hat{\rho}_q$,
applying the maximum entropy method (MEM) with
the optimum Langange multiplier (OLM) \cite{Martinez00} to the
generalized von Neumann (Tsallis) entropy given by Eq. (\ref{eq:A2})
under the constraints given by
\begin{eqnarray}
Tr \hat{\rho}_q &=& 1,\\
Tr \{ \hat{\rho}_q^q \hat{N} \} &=& Tr \hat{\rho}_q^q\: N_q, \nonumber \\
Tr \{ \hat{\rho}_q^q \hat{H} \} &=& Tr \hat{\rho}_q^q\: E_q, \nonumber
\end{eqnarray}
where 
$N_q$ and $E_q$ denote expectation values 
of the number operator ($\hat{N}$) and the Hamiltonian ($\hat{H}$), respectively.
The OLM-MEM leads to the density matrix given by \cite{Martinez00}
\begin{eqnarray}
\hat{\rho}_q &=& \frac{1}{X_q}\hat{w}, 
\label{eq:B12}
\end{eqnarray}
with
\begin{eqnarray}
X_q &=& Tr \:\hat{w}, 
\label{eq:B1}\\
N_q &=& \frac{1}{X_q} Tr\{\hat{w}^{q} \hat{N} \},
\label{eq:B2} \\
E_q &=& \frac{1}{X_q}Tr\{\hat{w}^{q} \hat{H} \}, 
\label{eq:B3}\\
\hat{w} &=& [1-(1-q)\beta(\hat{H}-\mu\hat{N}- E_q+\mu N_q)]^{1/(1-q)},
\label{eq:B4} 
\end{eqnarray}
where $\beta$ stands for the inverse temperature and $\mu$ 
the chemical potential (Fermi level).
The escort density matrix $\hat{P}_q$ is expressed by
\begin{eqnarray}
\hat{P}_q &=& \frac{\hat{\rho}_q^q}{c_q}
=\frac{1}{X_q} \hat{w}^{q},
\label{eq:B13}
\end{eqnarray}
where we have employed the relation,
\begin{eqnarray}
c_q &=& Tr \:\hat{\rho}_q^q = X_q^{1-q}.
\label{eq:B14}
\end{eqnarray}

By using the formulae for the gamma function $\Gamma(z)$ given by
\begin{eqnarray}
x^{-s} &=& \frac{1}{\Gamma(s)} 
\int_0^{\infty} u^{s-1} e^{-x u} \:du 
\hspace{1cm} \mbox{for $ \Re \:s > 0$},
\label{eq:B15}
\end{eqnarray}
\begin{eqnarray}
x^{s} &=& \frac{i}{2 \pi} \Gamma(s+1)\int_C (-t)^{-s-1} e^{-xt}\:dt
\hspace{1cm} \mbox{for $ \Re \:s > 0$},
\label{eq:B16}
\end{eqnarray}
we may express Eqs. (\ref{eq:B1})-(\ref{eq:B3}) as integrals along the
real axis for $q >1.0$ and in the complex plane for $q < 1.0$ 
as follows \cite{Rajagopal98,Lenzi99,Hasegawa09b}:
\begin{eqnarray}
X_q  = \left\{ \begin{array}{ll}
 \int_0^\infty  {G\left( {u;\frac{1}{{q - 1}},\frac{1}{{(q - 1)\beta }}} \right)
 \:Y_1(u)\:du \quad \mbox{for $q > 1$}, }  \\ 
 \frac{i}{{2\pi}}\int\limits_C {H\left( {t;\frac{1}{{1 - q}},\frac{1}{{(1 - q)\beta }}} \right) 
 Y_1(-t)\:dt \quad \mbox{for $q < 1$},}   
 \end{array} \right.
 \label{eq:B5}
\end{eqnarray}
\begin{eqnarray}
N_q  = \left\{ \begin{array}{ll}
 \frac{1}{{X_q }}\int_0^\infty  {G\left( {u;\frac{q}{{q - 1}},\frac{1}{{(q - 1)\beta }}} \right)
 Y_1(u)N_1(u) du \quad\mbox{for $q > 1$}, }  \\ 
 \frac{i}{{2\pi X_q }}\int\limits_C {H\left( {t;\frac{q}{{1 - q}},\frac{1}{{(1 - q)\beta }}} \right)
 Y_1(-t)N_1(-t) \:dt \quad \mbox{for $q < 1$},}   
 \end{array} \right.
 \label{eq:B6}
\end{eqnarray}
\begin{eqnarray}
E_q  = \left\{ \begin{array}{ll}
 \frac{1}{{X_q }}\int_0^\infty  {G\left( {u;\frac{q}{{q - 1}},\frac{1}{{(q - 1)\beta }}} \right)
 Y_1(u)E_1(u) \:du \quad\mbox{for $q > 1$}, }  \\ 
 \frac{i}{{2\pi X_q }}\int\limits_C {H\left( {t;\frac{q}{{1 - q}},\frac{1}{{(1 - q)\beta }}} \right)
 Y_1(-t)E_1(-t) \:dt \quad \mbox{for $q < 1$},}   
 \end{array} \right.
 \label{eq:B7}
\end{eqnarray}
where
\begin{eqnarray}
Y_1(u) &=& e^{u(E_q-\mu N_q)} \:\Xi_1(u), \\
\Xi_1(u) 
&=& Tr\: e^{-u(\hat{H}-\mu \hat{N})} 
=e^{ \mp \sum_k \ln[1 \mp e^{-u(\epsilon_k-\mu)}]}, 
\label{eq:B8} \\
%
N_1(u) &=& \sum_k \:f_1(\epsilon_k,u), \\
E_1(u) &=& \sum_k \:f_1(\epsilon_k,u) \:\epsilon_k, 
\label{eq:B9} \\
f_1(\epsilon,u) &=& \frac{1}{e^{u(\epsilon-\mu)} \mp 1},
\label{eq:B10} \\
G\left(u;a,b \right) 
&=& \frac{b^a}{\Gamma\left(a  \right)} 
u^{a-1} e^{-b u},
\label{eq:B11} \\
H(t;a,b)&=&  \: \Gamma(a+1) b^{-a} \:(-t)^{-a-1} e^{- b t}.
\end{eqnarray}
Upper and lower signs in Eqs. (\ref{eq:B8}) and (\ref{eq:B10})
are applied to boson and fermion, respectively,
and
$C$ denotes the Hankel path in the complex plane \cite{Rajagopal98,Lenzi99}.
Note that $N_q$ and $E_q$ are obtained by Eqs. (\ref{eq:B5})-(\ref{eq:B7}) 
in a self-consistent way. 
Such self-consistent calculations have been made 
for the electron band model and the Debye phonon model in \cite{Hasegawa09b}.

\subsection{Interpolation approach}

Self-consistent calculations including $N_q$ and $E_q$ 
are rather tedious and difficult.
In order to overcome this difficulty, we have proposed
the IA \cite{Hasegawa09b}, assuming 
that
\begin{eqnarray}
\frac{Y_1(u)}{X_q} &=&
\frac{1}{X_q} \:e^{u(E_q-\mu N_q)} \:\Xi_1(u) = 1,
 \label{eq:C0}
\end{eqnarray}
in Eqs. (\ref{eq:B6}) and (\ref{eq:B7}). Then they are expressed by
\begin{eqnarray}
N_q  = \left\{ \begin{array}{ll}
 \int_0^\infty  {G\left( {u;\frac{q}{{q - 1}},\frac{1}{{(q - 1)\beta }}} \right)
 N_1(u) \:du \quad\mbox{for $q > 1$}, }  \\ 
 \frac{i}{{2\pi}}\int\limits_C {H\left( {t;\frac{q}{{1 - q}},\frac{1}{{(1 - q)\beta }}} \right)
 N_1(-t) \:dt \quad \mbox{for $q < 1$},}   
 \end{array} \right.
 \label{eq:C1}
\end{eqnarray}
\begin{eqnarray}
E_q  = \left\{ \begin{array}{ll}
 \int_0^\infty  {G\left( {u;\frac{q}{{q - 1}},\frac{1}{{(q - 1)\beta }}} \right)
 E_1(u) \:du \quad\mbox{for $q > 1$}, }  \\ 
 \frac{i}{{2\pi}}\int\limits_C {H\left( {t;\frac{q}{{1 - q}},\frac{1}{{(1 - q)\beta }}} \right)
 E_1(-t) \:dt \quad \mbox{for $q < 1$}}.   
 \end{array} \right.
 \label{eq:C2}
\end{eqnarray}
Equations (\ref{eq:B6}) and (\ref{eq:B7}) are alternatively expressed by
\begin{eqnarray}
N_q &=& \sum_k f_q(\epsilon_k,\beta), \\
E_q &=& \sum_k f_q(\epsilon_k,\beta)\:\epsilon_k,
\end{eqnarray}
where the quantal distribution $f_q(\epsilon_k, \beta)$ is given by
\begin{eqnarray}
f_q (\epsilon_k ,\beta ) = \left\{ \begin{array}{ll}
 \int_0^\infty  {G\left( {u;\frac{q}{{q - 1}},\frac{1}{{(q - 1)\beta }}} \right)
 f_1 (\epsilon _k ,u)\:du \quad\mbox{for $q > 1$}, }  \\ 
 \frac{i}{{2\pi}}\int\limits_C {H\left( {t;\frac{q}{{1 - q}},\frac{1}{{(1 - q)\beta }}} \right)
 f_1 (\epsilon _k , - t)\:dt \quad \mbox{for $q < 1$}.}   
 \end{array} \right.
\label{eq:C3} 
\end{eqnarray}


With the use of Eq. (\ref{eq:C3}),
the analytic expression of the $q$-BED in the IA is given by \cite{Hasegawa09b}
\begin{eqnarray}
f_q(\epsilon,\beta) 
&=& \sum_{n=0}^{\infty}\:[e_q^{-(n+1)\:x}]^q
\hspace{1cm}\mbox{for $0 < q < 3 $}, 
\label{eq:C4}
\end{eqnarray}
where $x=\beta(\epsilon-\mu)$ and
$e_q^x$ expresses the $q$-exponential function defined by
\begin{eqnarray}
e_q^{x} = [1+(1-q)x]_+^{1/(1-q)},
\end{eqnarray}
with $[y]_+=y\:\Theta(y)$ and $\Theta(y)$ is the Heaviside function.


Similarly, the analytic expression of the $q$-FDD in the IA 
is given by \cite{Hasegawa09b}
\begin{eqnarray}
f_q(\epsilon, \beta) 
&=& \left\{ \begin{array}{ll}
F_q(x)
\quad & \mbox{for $ \epsilon  > \mu$}, \\
\frac{1}{2}
\quad & \mbox{for $ \epsilon  = \mu$}, \\
1.0 - F_q(\vert x \vert)
\quad & \mbox{for $ \epsilon < \mu $},
\label{eq:C5}
\end{array} \right.
\end{eqnarray}
with
\begin{eqnarray}
F_q(x) &=& \sum_{n=0}^{\infty} \: \: (-1)^n [e_q^{-(n+1) x}]^q
\hspace{1.cm}\mbox{for $0 < q < 3$},
\end{eqnarray}
where $x=\beta(\epsilon-\mu)$.
$f_q^{IA}(\epsilon, \beta)$ given by Eqs. (\ref{eq:C4}) and (\ref{eq:C5}) 
reduces to $f_1(\epsilon, \beta)$ in the limit of $q \rightarrow 1.0$
where $e_q^{x} \rightarrow e^{x}$. 

\section{Information Entropies}
\subsection{Tsallis entropy}

By using Eqs. (\ref{eq:A2}) and (\ref{eq:B14}), we obtain
\begin{eqnarray}
S_q &=& k_B \frac{1-c_q}{q-1} 
= k_B \ln_q (X_q),
\label{eq:D1}
\end{eqnarray}
where $X_q$ is given by Eq. (\ref{eq:B5}), or alternatively by
\begin{eqnarray}
X_q = \left\{ \begin{array}{ll}
\int_0^{\infty} 
G\left(u;\frac{1}{q-1}, \frac{1}{(q-1)\beta} \right) \: 
e^{u \sum_k (\epsilon_k-\mu) f_q(\epsilon_k,u) \pm 
\sum_k \ln[1 \pm f_1(\epsilon,u)]}
\:du, \quad \mbox{for $q>1$ }, \\
\frac{i}{2 \pi } 
\int_C H\left(t;\frac{1}{1-q},\frac{1}{(1-q)\beta} \right)
e^{-t \sum_k (\epsilon_k-\mu) f_q(\epsilon_k,-t) \pm 
\sum_k \ln[1 \pm f_1(\epsilon,-t)]}
\:dt \quad \mbox{for $q<1$ }, 
\end{array} \right. 
\label{eq:D2}
\end{eqnarray}
$\ln_q(x)$ being the $q$-logarithmic function,
\begin{eqnarray}
\ln_q(x)=\frac{x^{1-q}-1}{1-q}.
\label{eq:D3}
\end{eqnarray}

In the limit of $q \rightarrow 1.0$ where $\ln_q(x) \rightarrow \ln(x)$, 
Eq. (\ref{eq:D2}) reduces to
\begin{eqnarray}
X_1 &=& 
e^{\beta \sum_k (\epsilon_k-\mu) f_1(\epsilon_k,\beta) 
\pm \sum_k \ln[1 \pm f_1(\epsilon,\beta)]}, \nonumber
\end{eqnarray}
which yields the well-known expression of the quantum 
Boltzmann-Gibbs entropy given by
\begin{eqnarray}
S_1 &=& k_B \ln (X_1) 
= k_B \beta \left( \sum_k (\epsilon-\mu)f_1(\epsilon_k,\beta) 
\pm \sum_k [1\pm f_1(\epsilon_k,\beta)] \right), \nonumber \\
&=& -k_B \left( \sum_k f_1 \ln f_1 \mp \sum_k (1 \pm f_1) \ln (1 \pm f_1) \right).
\label{eq:D4}
\end{eqnarray}

\subsection{Generalized Fisher information matrix}

\subsubsection{$g_{\theta_n \theta_m}$}

The distance between two operators $\hat{\rho}$ and $\hat{\sigma}$
in the Riemann space is defined by
\begin{eqnarray}
D_q(\hat{\rho}\Vert \hat{\sigma})
&=& K_q(\hat{\rho} \Vert \hat{\sigma})+ K_q(\hat{\sigma} \Vert \hat{\rho}),
\label{eq:E5}
\end{eqnarray}
with the generalized Kullback-Leibler divergence $K_q(\hat{\rho} \Vert \hat{\sigma})$ 
\cite{Abe03b,Abe04},
\begin{eqnarray}
K_q(\hat{\rho} \Vert \hat{\sigma}) &=&
\frac{1}{1-q}(Tr \:\hat{\rho}-Tr \{ \hat{\rho}^q \hat{\sigma}^{1-q} \} ),
\label{eq:E1}
\end{eqnarray}
which is in conformity with the Tsallis entropy given by Eq. (\ref{eq:A2}).
The distance between the neighboring operators of 
$\hat{\rho}_q(\{\theta_n\})$ ($\equiv \hat{\rho}_q$)
and $\hat{\rho}_q(\{\theta_n+\delta \theta_n \})$ ($ \equiv \hat{\rho}'_q$)
is expressed by
\begin{eqnarray}
D_q(\hat{\rho}_q \Vert \hat{\rho}'_q) &\cong & 
\sum_{nm} \: g_{\theta_n \theta_m} \:\delta \theta_n \delta \theta_m,
\label{eq:E6}
\end{eqnarray}
where the GFI of $g_{\theta_n \theta_m}$ is given by \cite{Portesi07}
\begin{eqnarray}
g_{\theta_n \theta_m} &=&
q \left< \left( \frac{\partial \ln \hat{\rho}_q}{\partial \theta_n} \right)
\left( \frac{\partial \ln \hat{\rho}_q}{\partial \theta_m} \right) 
\right>_q,
\label{eq:E2}
\end{eqnarray}
the bracket $\langle \hat{Q} \rangle_q$ 
denoting the expectation value of an operator $\hat{Q}$ over $\hat{\rho}_q$,
\begin{eqnarray}
\langle \hat{Q} \rangle_q &=& Tr \; \{ \hat{\rho}_q \hat{Q} \}.
\end{eqnarray}
The GFI given by Eq. (\ref{eq:E2}) has a clear geometrical 
meaning expressing the metric between the adjacent density matrices 
in the Riemannian space spanned by the OLM density matrices of $\{\hat{\rho}_q\}$ 
\cite{Portesi07}.
Equations (\ref{eq:E1}) and (\ref{eq:E2}) are quantum extensions 
of the counterparts in the nonextensive classical statistics 
\cite{Masi06}.

For $(\theta_1, \theta_2)=(\beta, \beta\mu)$, Eqs. (\ref{eq:B12}) and (\ref{eq:E2}) 
yield
\begin{eqnarray}
g_{11} &=& \left( \frac{q}{X_q}\right)\:Tr\:\{ \hat{w}^{2q-1} (\hat{H}-E_q)^2\}, 
\label{eq:E3} \\
g_{22} &=& \left( \frac{q}{X_q}\right)\: Tr\: \{\hat{w}^{2q-1} (\hat{N}-N_q)^2 \}, \\
g_{12} &=& g_{21} 
=- \left( \frac{q}{X_q}\right)\: Tr\: \{ \hat{w}^{2q-1} ( \hat{H}-E_q)(\hat{N}-N_q) \}.
\label{eq:E4}
\end{eqnarray}

\subsubsection{$G_{\theta_n \theta_m}$}

Naudts proposed the quantum GFI given by \cite{Naudts05}
\begin{eqnarray}
G_{\theta_n \theta_m} 
&=& \left[ \left( \hat{P}_q^{-1} \frac{\partial \hat{\rho}_q}{\partial \theta_n} \right)
\left( \hat{P}_q^{-1} \frac{\partial \hat{\rho}_q}{\partial \theta_m} \right)
\right]_q,
\label{eq:F1}
\end{eqnarray}
where the bracket $[\hat{Q} ]_q$ denotes the expectation value given by
\begin{eqnarray}
\left[ \hat{Q} \right]_q &=& Tr \;\{\hat{P}_q \hat{Q}\}.
\end{eqnarray}
When we consider the expectation value of $[\hat{A} ]_q$
for an operator of $\hat{A}$ and $\theta_n=\theta_m=\theta$,
the generalized CRT is shown to be expressed by
\cite{Naudts05}
\begin{eqnarray}
[(\hat{A}-[\hat{A}]_q)^2]_q
\geq \frac{ (\frac{\partial A'}{\partial \theta})^2}
{G_{\theta \theta}},
\label{eq:F2}
\end{eqnarray}
where 
\begin{eqnarray}
A' &=& \langle \hat{A} \rangle_q.
\label{eq:F3}
\end{eqnarray}
Note that $A'$ is an average over $\hat{\rho}_q$. The CRT given by
Eq. (\ref{eq:F2}) may be derived as follows \cite{Naudts05}.
Taking the derivative of $A'$ with respect to $\theta$ and
using the relation: $\partial \hat{\rho}_q/\partial \theta=0$, we obtain
\begin{eqnarray}
\frac{\partial A'}{\partial \theta}
&=& Tr\{ \frac{\partial \hat{\rho}_q}{\partial \theta}(\hat{A}-[\hat{A}]_q)\}, \\
&=& Tr\{ \hat{P_q} \left(\hat{P_q}^{-1}\frac{\partial \hat{\rho}_q}{\partial \theta}\right)
(\hat{A}-[\hat{A}]_q)\}. 
\end{eqnarray}
Employing the Cauchy-Schwartz inequality, we obtain
\begin{eqnarray}
\left( \frac{\partial A'}{\partial \theta} \right)^2
&\leq& Tr\{ \hat{P_q} \left(\hat{P_q}^{-1}
\frac{\partial \hat{\rho}_q}{\partial \theta}\right)^2\}\;
Tr \{ \hat{P_q}(\hat{A}-[\hat{A}]_q)^2\},
\end{eqnarray}
which leads to Eq. (\ref{eq:F2}).

It is straightforward to extend the method mentioned above to the case of
$(\hat{A}_1, \hat{A}_2)=(\hat{H}, \hat{N})$
and $(\theta_1,\theta_2)=(\beta, \beta\mu)$.
The generalized CRT is expressed by
\begin{eqnarray}
\sf{V} \geq \sf{C}^{T} \:\sf{G}^{-1} \:\sf{C} \equiv \sf{D}, 
\label{eq:F4}
\end{eqnarray}
where the covariance matrix $\sf{V}$ is given by
\[\sf {V}=\left(
\begin{array}{cc}
\left[(\hat{H}-E_q)^2 \right]_q 
& \left[(\hat{H}-E_q)(\hat{N}-N_q) \right]_q \\
\left[(\hat{H}-E_q)(\hat{N}-N_q) \right]_q & 
\left[(\hat{N}-N_q)^2 \right]_q \\
\end{array}
\right). \]
Calculations using Eqs. (\ref{eq:B13}) and (\ref{eq:F1}) yield
elements of the Fisher information matrix $\sf{G}$ given by
\begin{eqnarray}
G_{11} &=& \left[ (\hat{H}-E_q)^2 \right]_q = V_{11}, 
\label{eq:F5} \\
G_{22} &=& \left[(\hat{N}-N_q)^2 \right]_q= V_{22}, 
\label{eq:F6} \\
G_{12} &=& G_{21}
= - \: \left[ (\hat{H}-E_q)(\hat{N}-N_q) \right]_q = - V_{12}=-V_{21},
\label{eq:F7} 
\end{eqnarray}
and those of $\sf{C}$ expressed by
\begin{eqnarray}
C_{11} &=& 
\frac{\partial \langle \hat{H} \rangle_q }{\partial \beta} 
= -\left[ \hat{H}(\hat{H}-E_q)  \right]_q, 
\label{eq:F8}\\
C_{22} &=& \frac{\partial \langle \hat{N} \rangle_q}{\partial (\beta\mu)}
= \left[ \hat{N}(\hat{N}-N_q)  \right]_q, \\
C_{12} &=& \frac{\partial \langle \hat{H} \rangle_q}{\partial (\beta\mu)}
= \left[ \hat{H}(\hat{N}-N_q)  \right]_q, \\
C_{21} &=& \frac{\partial \langle \hat{N} \rangle_q}{\partial \beta} 
= - \left[ \hat{N}(\hat{H}-E_q)  \right]_q.
\label{eq:F9}
\end{eqnarray}
A simple calculation leads to
\begin{eqnarray}
 \sf{D} &=&\sf{V},
\label{eq:F10}
\end{eqnarray}
which implies that the CRT given by Eq. (\ref{eq:F4}) is
satisfied with an equal sign.

\subsubsection{$\tilde{g}_{\theta_n \theta_m}$}

The GFI given by
\begin{eqnarray}
\tilde{g}_{\theta_n \theta_m} &=&
\left[ \left( \frac{\partial \ln \hat{P}_q}{\partial \theta_n} \right)
\left( \frac{\partial \ln \hat{P}_q}{\partial \theta_m} \right) \right]_q,
\label{eq:J0} 
\end{eqnarray}
provides the lower bound of unbiased estimates in the CRT as shown in the following.
For the expectation value of $A=[\hat{A}]_q$, we obtain
\begin{eqnarray}
\frac{\partial A}{\partial \theta} 
&=& Tr \{\frac{\partial \hat{P}_q}{\partial \theta}(\hat{A}-[\hat{A}]_q)\}, \\
&=& Tr \{\hat{P}_q \left( \hat{P}_q^{-1} 
\frac{\partial \hat{P}_q}{\partial \theta} \right)(\hat{A}-[\hat{A}]_q) \}.
\end{eqnarray}
By using the Cauchy-Schwartz inequality, we then obtain the CRT given by
\begin{eqnarray}
[(\hat{A}-[\hat{A}]_q)^2]_q
\geq \frac{ (\frac{\partial A}{\partial \theta})^2}
{\tilde{g}_{\theta \theta}}.
\label{eq:J8}
\end{eqnarray}
With the use of Eq. (\ref{eq:B13}) and (\ref{eq:J0}), the generalized CRT
for the case of $(\hat{A}_1, \hat{A}_2)=(\hat{H}, \hat{N})$
and $(\theta_1, \theta_2)=(\beta, \beta\mu)$ 
is expressed by
\begin{eqnarray}
\sf{V} \geq \tilde{\sf{C}}^{T} \:\tilde{\sf{g}}^{-1} \:\tilde{\sf{C}}
\equiv \tilde{\sf{D}},
\label{eq:J1}
\end{eqnarray}
where elements of $\tilde{\sf{g}}$ are given by
\begin{eqnarray}
\tilde{g}_{11} &=& \left( \frac{q^2}{X_q} \right)
\: Tr\: \{\hat{w}^{3q-2} (\hat{H}-E_q)^2 \}, 
\label{eq:J2}\\
\tilde{g}_{22} &=& \left( \frac{q^2}{X_q} \right)
\: Tr\: \{\hat{w}^{3q-2} (\hat{N}-N_q)^2 \}, 
\label{eq:J3}\\
\tilde{g}_{12} &=& \tilde{g}_{21} = - \left( \frac{q^2}{X_q} \right)
\: Tr\: \{ \hat{w}^{3q-2} (\hat{H}-E_q)(\hat{N}-N_q) \}, 
\label{eq:J4}
\end{eqnarray}
and those of $\tilde{\sf{C}}$ are given by
\begin{eqnarray}
\tilde{C}_{11} &=& \frac{\partial E_q}{\partial \beta} 
= -\left( \frac{q}{X_q}\right) 
Tr \:\{\hat{w}^{2q-1} 
\hat{H}(\hat{H}-E_q) \}, 
\label{eq:J5}\\
\tilde{C}_{22} &=& \frac{\partial N_q}{\partial (\beta\mu)} 
= \left( \frac{q}{X_q}\right) 
Tr \:\{\hat{w}^{2q-1} 
\hat{N}(\hat{N}-N_q) \}, \\
\tilde{C}_{12} &=& \frac{\partial E_q}{\partial (\beta\mu)} 
= \left( \frac{q}{X_q}\right) 
Tr \:\{\hat{w}^{2q-1} 
\hat{H}(\hat{N}-N_q) \}, \\
\tilde{C}_{21} &=& \frac{\partial N_q}{\partial \beta} 
= -\left( \frac{q}{X_q}\right) 
Tr \:\{\hat{w}^{2q-1} 
\hat{N}(\hat{H}-E_q) \}.
\label{eq:J6}
\end{eqnarray}
It is noted note that the GFI given by Eq. (\ref{eq:J0}) is nothing but
a quantum extension of that proposed in nonextensive classical
statistics \cite{Chimento00,Pennini04,Hasegawa08}.

We may show that $\tilde{g}_{\theta_n \theta_m}$ given by Eq. (\ref{eq:J0}) 
denotes the metric tensor for the neighboring escort density matrices
of $\hat{P}_q(\{\theta_n \})$ $(\equiv \hat{P}_q )$ 
and $\hat{P}_q(\{\theta_n+ \delta \theta_n \})$ $(\equiv \hat{P}'_q )$
as given by
\begin{eqnarray}
D_1(\hat{P}_q \Vert \hat{P}'_q) &\cong & 
\sum_{nm} \:\tilde{g}_{\theta_n \theta_m} \:\delta \theta_n \delta \theta_m,
\label{eq:J7}
\end{eqnarray}
where $D_1(\hat{\rho} \Vert \hat{\sigma})$ stands for the distance given 
by Eqs. (\ref{eq:E5}) and (\ref{eq:E1}) with $q=1.0$.

\subsubsection{A comparison among $\sf{g}$, $\sf{G}$ and $\tilde{\sf{g}}$}

In the preceding subsections, 
we have  discussed the properties of $\sf{g}$, $\sf{G}$ 
and $\tilde{\sf{g}}$. A comparison among the three GFIs is made in Table 1.
We note that $\sf{g}$ expresses the metric in the Riemann space
spanned by density matrices ($\hat{\rho}_q$), but $\sf{G}$ does not. 
In contrast, $\sf{G}$ provides the lower bound for the CRT, to which
$\sf{g}$ is not applicable. On the contrary,
$\tilde{\sf{g}}$ denotes the distance between the escort density
matrices ($\hat{P}_q$) and it also satisfies the CRT.
It is shown that $\sf{G}$ provides a better bound for the inequality 
in the CRT than $\tilde{\sf{g}}$: $\sf{V}=\sf{D} \geq \tilde{\sf{D}}$.

\begin{table}
\begin{center}
\caption{A comparison among three GFIs for
$\theta_n=\theta_m=\theta$ ($q \neq 1.0$)}
\renewcommand{\arraystretch}{1.5}
\begin{tabular}{|c|c|c|c|} \hline

GFI &  \hspace{0.4cm}Metric \hspace{0.4cm} & \hspace{0.5cm}CRT\hspace{0.5cm}  
& \hspace{0.4cm} Refs.\hspace{0.4cm} \\ \hline \hline
$g_{\theta \theta}=q \:Tr\{\hat{\rho}_q^{-1} 
(\partial \hat{\rho}_q/\partial \theta )^2\}$  
& Eq. (\ref{eq:E6})  
& NA
& \cite{Masi06,Portesi07}  \\ \hline
$G_{\theta \theta}=Tr \{\hat{P}_q^{-1}
(\partial \hat{\rho}_q/\partial \theta)^2\}$  
& NA  &  Eq. (\ref{eq:F2})
& \cite{Naudts04,Naudts05}  \\ \hline
$\tilde{g}_{\theta\theta}= Tr \{\hat{P}_q^{-1} 
(\partial \hat{P}_q/\partial \theta)^2\}$ 
&  Eq. (\ref{eq:J7})
&  Eq. (\ref{eq:J8})
&  \cite{Chimento00,Pennini04,Hasegawa08} \\ \hline

\end{tabular}
\end{center}

CRT: Cram\'{e}r-Rao theorem,
NA: not applicable

\end{table}

When we adopt the exact transformation with the use of formulae for the gamma 
function given by Eqs. (\ref{eq:B15}) and (\ref{eq:B16}), 
equations for the GFIs may be expressed as integrals
along real axis for $q \geq 1.0$ or in the complex plane for $q < 1.0$.
Expressions of the GFIs for $q \geq 1.0$ with the IA [Eq. (\ref{eq:C0})] 
are summarized in the Appendix.
It is now possible for us to numerically calculate the Tsallis entropy $S_q$
and the GFI matrices
as functions of $q$ and temperature. We will report such numerical calculations
for the electron band model and the Debye phonon model in the
following section.

\section{Numerical calculations}
\subsection{Electron band model}
We employ a band model for electrons with 
a uniform density of state given by \cite{Hasegawa09b}
\begin{eqnarray}
\rho(\epsilon)=(1/2 W) \;\Theta(W- \vert \epsilon \vert),
\label{eq:G1}
\end{eqnarray}
where $W$ denotes a half of the total band width.
We adopt $N=0.5$, for which $\mu=0.0$ independently of 
the temperature because of the adopted symmetric density of states given 
by Eq. (\ref{eq:G1}).

Figure \ref{fig1} shows the temperature dependence of $E_q$ for various
$q$ calculated self-consistently with the use of Eqs. (\ref{eq:B5})-(\ref{eq:B7}). 
With increasing $q$ from unity, $E_q$ at higher temperatures
is decreased, although that at lower temperatures is increased
(see also the inset of Fig. 1 of Ref. \cite{Hasegawa08}).
We have calculated the average energy also by using Eq. (\ref{eq:C2}) in the IA. 
The ratio of $E_q^{IA}/E_q$ ($\equiv \lambda$) is plotted in the inset of Fig. \ref{fig1}.
For $q=1.5$, for example, this ratio is changed from unity at low
temperatures ($k_B T/W \sim 0.005$) 
to about 0.991 at high temperatures ($k_B T/W \sim 1.0$).
For $1.0 < q < 1.5$, an agreement between $E_q$ and $E_q^{(IA)}$
is much better.

The temperature dependence of $S_q$ for various $q$ is shown in Fig. \ref{fig2}.
With increasing $q$ from unity, the temperature dependence of
$S_q$ at low temperatures becomes more significant but 
its saturated value at high temperatures become smaller.

Solid, chain and dotted curves in Fig. \ref{fig3} show
temperature dependence of $G_{ii}$, $g_{ii}$ 
and $\tilde{g}_{ii}$ ($i=1,2$), respectively, for $q=1.1$:
results for $q=1.0$ are plotted by dashed curves.
Note that $V_{ii}=G_{ii}$ [Eqs. (\ref{eq:F5}) and (\ref{eq:F6})] 
which implies that the temperature dependence
of the variance $V_{ii}$ is the same as that of $G_{ii}$.
All the GFIs show a similar temperature dependence.
A closer inspection, however, shows that there are some differences
between them. In particular, with increasing $q$ from unity,
$g_{ii}$ and $\tilde{g}_{ii}$ are increased while $G_{ii}$ is decreased.

\subsection{Debye phonon model}

We adopt the Debye model whose phonon density of states
is given by \cite{Hasegawa09b}
\begin{eqnarray}
\rho(\omega) &=& A \: \omega^2 
\hspace{1cm} \mbox{for $0 < \omega \leq \omega_D$},
\label{eq:H1}
\end{eqnarray}
where $A= 9 N_a/w_D^3$, $N_a$ denotes the number
of atoms, $\omega$ the phonon frequency
and $\omega_D$ the Debye cutoff frequency.

Figure \ref{fig4} shows the temperature dependence of $E_q$
for various $q$ (with $\mu=0.0$), which is self-consistently calculated
by Eqs. (\ref{eq:B5})-(\ref{eq:B7}).
It shows that $E_q$ is larger for larger $q$.
The inset of Fig. \ref{fig4} shows the ratio of $E_q^{IA}/E_q$ ($=\lambda$)
where $E_q^{IA}$ is calculated within the IA [Eq.(\ref{eq:C2})].
Although the ratio is not good at $0.02 \lesssim T/T_D \lesssim 0.5$,
it becomes better at $1 \lesssim T/T_D \lesssim 2$, 
where $T_D$ signifies the Debye temperature ($k_B T_D=\hbar \omega_D$).

Figure \ref{fig5} shows the temperature dependence of $S_q$.
We note that for larger $q$, the temperature dependence
at low temperatures becomes more steep and its saturated value 
at high temperatures
becomes smaller.

Solid, chain and dotted curves in Fig. \ref{fig6} show
the temperature dependence of $G_{11}$, $g_{11}$
and $\tilde{g}_{11}$, respectively, for $q=1.1$: 
results for $q=1.0$ are shown by dashed curves.
Note that the temperature dependence of the variance $V_{11}$
is the same as that of $G_{11}$ [Eq. (\ref{eq:F5})].
All the GFIs show a similar temperature dependence, although
with increasing $q$, magnitudes of $G_{ii}$ and $g_{ii}$ becomes 
larger while that of $\tilde{g}_{ii}$ becomes smaller.

Filled and open marks in Fig. \ref{fig7} show the $q$ dependence of
$\sf{D}$ and $\tilde{\sf{D}}$, respectively, for the Debye phonon model:
note that $\sf{D}=\sf{V}$ for the GFI in \cite{Naudts05}.
It is shown that the ratio of $\tilde{\sf{D}}/\sf{V}$, which is unity
for $q=1.0$, is gradually reduced with increasing $q$.

\section{Conclusion and discussion}

It is instructive to make a comparison among $\sf{g}$, $\sf{G}$
and $\tilde{\sf{g}}$ for the $q$-Gaussian distribution
in classical nonextensive statistics.
For given mean ($\mu_q$) and variance ($\sigma_q^2$), 
the $q$-Gaussian distribution
$p_q(x)$ and its escort distribution $P_q(x)$ are expressed by \cite{Hasegawa08}
\begin{eqnarray}
p_q(x) &=& \frac{1}{Z_q} 
\left[1-(1-q)\frac{(x-\mu_q)^2}{2 \nu \sigma_q^2} \right]^{1/(1-q)}, 
\label{eq:K1} \\
P_q(x) &=& \frac{1}{\nu Z_q} 
\left[1-(1-q)\frac{(x-\mu_q)^2}{2 \nu \sigma_q^2} \right]^{q/(1-q)},
\label{eq:K2}
\end{eqnarray}
with
\begin{eqnarray}
Z_q 
&=& \left\{ \begin{array}{ll}
\left(\frac{2 \nu \sigma_q^2}{q-1} \right)^{1/2}B
\left(\frac{1}{2}, \frac{1}{q-1}-\frac{1}{2} \right)
&\quad\mbox{for $1< q < 3$}, 
\\
\sqrt{2 \pi} \sigma_q
&\quad\mbox{for $q=1$}, 
\\
\left(\frac{2 \nu \sigma_q^2}{1-q} \right)^{1/2}
B\left(\frac{1}{2}, \frac{1}{1-q}+1 \right)
&\quad\mbox{for $0< q < 1$},
\end{array} \right. 
\label{eq:K3} 
\end{eqnarray}
where $\nu=(3-q)/2$ and $B(a,b)$ stands for the Beta function.
The GFIs given by Eqs. (\ref{eq:E2}), (\ref{eq:F1}) and (\ref{eq:J0}) 
in the classical case are expressed by
\begin{eqnarray}
g_{ij} &=& q \int p_q(x) 
\left(\frac{1}{p_q(x)} \frac{\partial p_q(x)}{\partial \theta_i} \right)
\left(\frac{1}{p_q(x)} \frac{\partial p_q(x)}{\partial \theta_j} \right) \:dx, 
\label{eq:K4} \\
G_{ij} &=& \int P_q(x) 
\left(\frac{1}{P_q(x)} \frac{\partial p_q(x)}{\partial \theta_i} \right)
\left( \frac{1}{P_q(x)} \frac{\partial p_q(x)}{\partial \theta_j} \right) \:dx, 
\label{eq:K5} \\
\tilde{g}_{ij} &=& \int P_q(x) 
\left(\frac{1}{P_q(x)} \frac{\partial P_q(x)}{\partial \theta_i} \right)
\left( \frac{1}{P_q(x)}\frac{\partial P_q(x)}{\partial \theta_j} \right) \:dx.
\label{eq:K6}
\end{eqnarray}
Averages over $p_q(x)$ and $P_q(x)$ are expressed by 
$\langle \hat{Q} \rangle_q$ and $[\hat{Q}]_q$, respectively.

For $(A_1, A_2)=([ x ]_q, [ (\delta x)^2]_q)$ and 
$(\theta_1, \theta_2)=(\mu_q, \sigma_q^2)$ where $\delta x=x-\mu_q$, 
we obtain \cite{Hasegawa08}

\[
\sf{g}=\left(
\begin{array}{cc}
\frac{1}{\sigma_q^2} & 0\\
0 & \frac{(3-q)}{4 \sigma_q^4}  \\
\end{array}
\right),
\]

\[
\sf \tilde{g}=\left(
\begin{array}{cc}
\frac{q(q+1)}{(3-q)(2 q-1)\sigma_q^2} & 0\\
0 & \frac{(q+1)}{4(2q-1)\sigma_q^4}  \\
\end{array}
\right),
\]

\[
\sf{V}= \left(
\begin{array}{cc}
\sigma_q^2 & 0 \\
0 & \frac{4 \sigma_q^4}{(5-3 q)} \\
\end{array}
\right),
\] 

\[\tilde{{\sf C}}= \left(
\begin{array}{cc}
\frac{\partial [ x ]_q}{\partial \mu_q} &
\frac{\partial [ x ]_q}{\partial \sigma_q^2}  \\
\frac{\partial [ (\delta x)^2 ]_q}{\partial \mu_q} & 
\frac{\partial [ (\delta x)^2 ]_q}{\partial \sigma_q^2} \\
\end{array}
\right)
= \left(
\begin{array}{cc}
1 & 0 \\
0 & 1 \\
\end{array}
\right).
\]
The CRT is then expressed by
\[
\sf{V} \geq \tilde{\sf{C}}^{T} \sf \tilde{\sf{g}}^{-1} \tilde{\sf{C}}
\equiv \sf{\tilde{D}}=  \left(
\begin{array}{cc}
\frac{(3-q)(2 q-1)\sigma_q^2}{q(q+1)} & 0\\
0 & \frac{4(2q-1)\sigma_q^4}{(q+1)}  \\
\end{array}
\right).
\]

On the contrary, for the GFI of $\sf{G}$ \cite{Naudts04,Naudts05} 
with $(A'_1, A'_2)=(\langle x \rangle_q, \langle (\delta x)^2 \rangle_q) $ and 
$(\theta_1, \theta_2)=(\mu_q, \sigma_q^2)$, we obtain 

\[
\sf G =\left(
\begin{array}{cc}
\frac{1}{\sigma_q^2} & 0\\
0 & \frac{(3-q)^2}{4(5-3q)\sigma_q^4}  \\
\end{array}
\right),
\]

\[{\sf C}= \left(
\begin{array}{cc}
\frac{\partial \langle x \rangle_q}{\partial \mu_q} &
\frac{\partial \langle x \rangle_q}{\partial \sigma_q^2}  \\
\frac{\partial \langle (\delta x)^2 \rangle_q}{\partial \mu_q} & 
\frac{\partial \langle (\delta x)^2 \rangle_q}{\partial \sigma_q^2} \\
\end{array}
\right)
= \left(
\begin{array}{cc}
1 & 0 \\
0 & \frac{3-q}{5-3 q} \\
\end{array}
\right).
\]
The CRT is expressed by
\[
\sf{V} \geq \sf{C}^{T} \sf{G}^{-1} \sf{C} 
\equiv \sf{D}= \left(
\begin{array}{cc}
\sigma_q^2 & 0 \\
0 & \frac{4 \sigma_q^4}{(5-3 q)} \\
\end{array}
\right),
\] 
which shows that the CRT is satisfied with an equal sign.

Solid curves in Figs \ref{fig8}(a) and \ref{fig8}(b) show 
$D_{11}/\sigma_q^2$ and $D_{22}/\sigma_q^4$, respectively,
dotted curves expressing $V_{11}/\sigma_q^2$ and $V_{22}/\sigma_q^4$.
Chain curves in  Figs \ref{fig8}(a) and \ref{fig8}(b) show 
$\tilde{D}_{11}/\sigma_q^2$ and $\tilde{D}_{22}/\sigma_q^4$, respectively.
For a comparison, we show by the dashed curve in Fig. \ref{fig8}(b), 
$1/g_{22}\sigma_q^4$ \cite{Hasegawa08}.
These figures clearly show that 
$\sf{G}$ and $\tilde{\sf{g}}$ preserve the CRT
and that $\sf{g}$ is not applicable to the CRT
because $V_{22} < 1/g_{22}$ for $q < 1$ in Fig. \ref{fig8}(b). 
The $q$ dependence of $\sf{\tilde{\sf{D}}}/\sf{V}$ 
in Figs. \ref{fig8}(a) and \ref{fig8}(b)
is similar to that for Debye phonon model shown in Fig. \ref{fig7}.

Adopting the factorization approximation (FA)
to a calculation of the grand-canonical partition function,
B\"{u}y\"{u}kkilic, Demirhan and G\"{u}lec \cite{Buy95} 
derived the generalized quantal distribution given by
\begin{eqnarray}
f_q^{FA}(\epsilon) 
&=& \frac{1}{ \{e_q[-\beta(\epsilon-\mu)] \}^{-1}\mp 1 }, 
\label{eq:L1}
\end{eqnarray}
where the upper (lower) sign is applied to $q$-BED ($q$-FDD).
It has been shown that $f_q^{FA}(\epsilon_k)$
may be alternatively derived by applying the variational condition to 
the entropy $S_q^{FA}$ given by \cite{Tewel05}
\begin{eqnarray}
S_q^{FA} &=& 
- \sum_k \{ [f_q^{FA}(\epsilon_k)]^q \ln_q[f_q^{FA}(\epsilon_k)]
\mp [ 1\pm f_q^{FA}(\epsilon_k)]^q \ln_q[1 \pm f_q^{FA}(\epsilon_k)] \}, 
\label{eq:L2}
\end{eqnarray}
with the constraints:
\begin{eqnarray}
\sum_k [f_q^{FA}(\epsilon_k)]^q &=& N, \nonumber \\
\sum_k [f_q^{FA}(\epsilon_k)]^q \;\epsilon_k &=& E. \nonumber
\end{eqnarray}
Quite recently the $q$-BED and $q$-FDD given by Eq. (\ref{eq:L1}) 
in the FA are criticized based on the exact approach \cite{Hasegawa09b}.
It has been pointed out that
the $O(q-1)$-order contribution in the FA does not agree
with that of the exact approach
and that its $q$-FDD yields inappropriate results even
qualitatively.
This criticism is applied also to the expression for
the FA entropy given by Eq. (\ref{eq:L2}).

To summarize,
we have discussed the generalized von Neumann (Tsallis) entropy
and the GFI in nonextensive quantum systems,
by using the IA \cite{Hasegawa09b}.
Numerical calculations of the $q$- and temperature-dependent
information entropies have been performed for the electron band model 
and the Debye phonon model.
A comparison among the three GFIs (Table 1) has shown that for the CRT 
in the nonextensive statistics, we have to employ $\sf{G}$ \cite{Naudts05} 
or $\tilde{\sf{g}}$ \cite{Chimento00,Pennini04,Hasegawa08} rather than $\sf{g}$ 
\cite{Portesi07} which has a geometrical meaning derived 
from the generalized Kullback-Leibler divergence.
Although our present discussion has been confined to $q \geq 1.0$ 
for nonextensive quantum systems, it is necessary to extend our study 
to the case of $q < 1.0$, which is our future subject.

\section*{Acknowledgments}
This work is partly supported by
a Grant-in-Aid for Scientific Research from the Japanese 
Ministry of Education, Culture, Sports, Science and Technology.  

\appendix

\section{The generalized Fisher information in the IA} 
\renewcommand{\theequation}{A\arabic{equation}}
\setcounter{equation}{0}

By using formulae for the gamma function given by Eqs. (\ref{eq:B15}) 
and (\ref{eq:B16}) and adopting the IA given by Eq. (\ref{eq:C0}),
we obtain following expressions for the GFIs.

Elements of $\sf{g}$ in Eqs. (\ref{eq:E3})-(\ref{eq:E4}) for $q \geq 1.0$ 
are expressed by
\begin{eqnarray}
g_{11} 
&=& q \:\int_0^{\infty} G\left(u;\frac{2q-1}{q-1}, \frac{1}{(q-1)\beta} \right) 
\left[ E_1^{(2)}(u) -2 E_q E_1(u)+ E_q^2 \right]\:du, \nonumber \\
&& \\
g_{22} 
&=& q\:\int_0^{\infty} G\left(u;\frac{2q-1}{q-1}, \frac{1}{(q-1)\beta} \right) 
\left[ N_1^{(2)}(u) -2 N_q N_1(u)+ N_q^2 \right]\:du, \nonumber \\
&& \\
g_{12} &=& g_{21} 
= -q \int_0^{\infty} G\left(u;\frac{2q-1}{q-1}, \frac{1}{(q-1)\beta} \right) \nonumber \\
&& \times
\left[ D_1^{(2)}(u) - E_q N_1(u) -N_q E_1(u)+E_q N_q \right]\:du, 
\end{eqnarray}
with
\begin{eqnarray}
N_1^{(2)}(u) &=& [N_1(u)]^2
+\sum_k f_1(\epsilon_k,u)[1 \pm f_1(\epsilon_k,u) ], \\
%
%
E_1^{(2)}(u) 
&=& [E_1(u)]^2
+\sum_k f_1(\epsilon_k,u) [1 \pm f_1(\epsilon_k,u) ]\: \epsilon_k^2, \\
D_1^{(2)}(u) &=&E_1(u)N_1(u) 
+\sum_k f_1(\epsilon_k,u)[1 \pm f_1(\epsilon_k,u) ] \:\epsilon_k, 
\end{eqnarray}
where 
the upper (lower) sign is applied to boson (fermion).
Relevant results for $q < 1.0$ may be obtainable
with a proper modification.

Elements of matrix of $\sf{G}$  
in Eqs. (\ref{eq:F5})-(\ref{eq:F7}) are expressed by
\begin{eqnarray}
G_{11} &=& 
\left[\hat{H}^2\right]_q-E_q^2, \\
G_{22} &=& 
\left[\hat{N}^2 \right]_q-N_q^2, \\
G_{12} &=& G_{21} 
= - \: \left(\left[ \hat{H} \hat{N} \right]_q-E_q N_q \right), 
\end{eqnarray}
with
\begin{eqnarray}
\left[\hat{H}^2 \right]_q
&=& \int_0^{\infty} G\left(u;\frac{q}{q-1}, \frac{1}{(q-1)\beta} \right) 
E_1^{(2)}(u)\:du, \\ 
\left[\hat{N}^2 \right]_q
&=& \int_0^{\infty} G\left(u;\frac{q}{q-1}, \frac{1}{(q-1)\beta} \right) 
N_1^{(2)}(u)\:du, \\ 
\left[\hat{H}\hat{N} \right]_q
&=& \int_0^{\infty} G\left(u;\frac{q}{q-1}, \frac{1}{(q-1)\beta} \right) 
D_1^{(2)}(u)\:du.
\end{eqnarray}

Elements of matrix of $\tilde{\sf{g}}$  
in Eqs. (\ref{eq:J2})-(\ref{eq:J4}) are expressed by
\begin{eqnarray}
\tilde{g}_{11} 
&=& q \:\int_0^{\infty} G\left(u;\frac{3q-2}{q-1}, \frac{1}{(q-1)\beta} \right) 
\left[ E^{(2)}(u) -2 E_q E_1(u)+ E_q^2 \right]\:du, \nonumber \\
&& \\
\tilde{g}_{22} 
&=& q\:\int_0^{\infty} G\left(u;\frac{3q-2}{q-1}, \frac{1}{(q-1)\beta} \right) 
\left[ N^{(2)}(u) -2 N_q N_1(u)+ N_q^2 \right]\:du, \nonumber \\
&& \\
\tilde{g}_{12} &=& g_{21} 
= -q \int_0^{\infty} G\left(u;\frac{3q-2}{q-1}, \frac{1}{(q-1)\beta} \right) \nonumber \\
&& \times
\left[ D^{(2)}(u) - E_q N_1(u) -N_q E_1(u)+E_q N_q \right]\:du. 
\end{eqnarray}

In the limit of $q \rightarrow 1.0$, all the GFIs reduce to
\begin{eqnarray}
g_{11} &=& G_{11} = \tilde{g}_{11}
=\frac{1}{\langle \hat{H}^2 \rangle_1 - \langle \hat{H} \rangle_1^2},\\
g_{22} &=& G_{22} = \tilde{g}_{22}
=\frac{1}{\langle \hat{N}^2 \rangle_1 - \langle \hat{N} \rangle_1^2},\\
g_{12} &=& G_{21} = \tilde{g}_{12}
=\frac{1}{\langle \hat{H}\hat{N} \rangle_1 
- \langle \hat{H} \rangle_1 \langle \hat{N} \rangle}_1.
\end{eqnarray}


\newpage

\begin{figure}
\begin{center}
\end{center}
\caption{
(Color online)
The temperature dependence 
of $E_q$ of the electron band model for $q=1.0$ (dashed curves), 
$q=1.1$ (chain curves), $q=1.2$ (dotted curves), 
$q=1.3$ (solid curves) and $q=1.5$ (double-chain curve),
the inset showing the ratio of $\lambda=E_q^{IA}/E_q$. 
}
\label{fig1}
\end{figure}

\begin{figure}
\begin{center}
\end{center}
\caption{
(Color online)
The temperature dependence 
of the Tsallis entropy $S_q$ of the electron band model 
for $q=1.0$ (the dashed curve), 
$q=1.1$ (the chain curve), $q=1.2$ (the dotted curve), 
$q=1.3$ (the solid curve) and $q=1.5$ (the double-chain curve).
}
\label{fig2}
\end{figure}

\begin{figure}
\begin{center}
\end{center}
\caption{
(Color online)
The temperature dependence of the $(i,i)$ component ($i=1,2$) 
of $\sf{G}$ (solid curves), $\sf{g}$ (chain curves) 
and $\tilde{\sf{g}}$ (dotted curves) for $q=1.1$ of the electron band model,
results for $q=1.0$ being plotted by dotted curves for a comparison. 
}
\label{fig3}
\end{figure}

\begin{figure}
\begin{center}
\end{center}
\caption{
(Color online)
The temperature dependence 
of $E_q$ of the Debye phonon model for $q=1.0$ (dashed curves), 
$q=1.1$ (chain curves), $q=1.2$ (dotted curves), $q=1.3$ (solid curves)
and $q=1.5$ (double-chain curves),
the inset showing the ratio of $\lambda = E_q^{IA}/E_q$. 
}
\label{fig4}
\end{figure}

\begin{figure}
\begin{center}
\end{center}
\caption{
(Color online)
The temperature dependence 
of the Tsallis entropy $S_q$ of the Debye phonon model
for $q=1.0$ (the dashed curve), 
$q=1.1$ (the chain curve), $q=1.2$ (the dotted curve), 
$q=1.3$ (the solid curve) and $q=1.5$ (the double-chain curve).
}
\label{fig5}
\end{figure}

\begin{figure}
\begin{center}
\end{center}
\caption{
(Color online)
The temperature dependence of the $(i,i)$ component ($i=1,2$) 
of $\sf{G}$ (solid curves), $\sf{g}$ (chain curves) 
and $\tilde{\sf{g}}$ (dotted curves) for $q=1.1$ of the Debye phonon model,
results for $q=1.0$ being plotted by dashed curves for a comparison. 
}
\label{fig6}
\end{figure}

\begin{figure}
\begin{center}
\end{center}
\caption{
(Color online)
The $q$ dependence of 
$D_{11}$ (filled circles), 
$D_{22}$ (filled squares),
$D_{12}$ (filled triangles),
$\tilde{D}_{11}$ (open circles),
$\tilde{D}_{22}$ (open squares) and
$\tilde{D}_{12}$ (open triangles) with $T/T_D=1.0$ 
for the Debye phonon model:
note that $V_{ij}=D_{ij}$ ($i,j=1,2$).
}
\label{fig7}
\end{figure}

\begin{figure}
\begin{center}
\end{center}
\caption{
(Color online)
The $q$ dependence of 
(a) $V_{11}/\sigma_q^2$ (the dotted curve), 
$D_{11}/\sigma_q^2$  (the solid curve),
$\tilde{D}_{11}/ \sigma_q^2$ (the chain curve)
and $ (1/g_{11} \sigma_q^2)$ (the dashed curve),
and (b) that of $V_{22}/\sigma_q^4$ (the dotted curve), 
$D_{22}/\sigma_q^4$ (the solid curve),
$\tilde{D}_{22}/\sigma_q^4$ (the chain curve) and 
$1/g_{22} \sigma_q^4$ (the dashed curve):
note that $V_{11}=D_{11}=1/g_{11}$ in (a) and 
$V_{22}=D_{22}$ in (b).
}
\label{fig8}
\end{figure}

\end{document}